\documentclass[a4paper,11pt]{article}
\usepackage{pos}

\usepackage{array} % array
\usepackage{bm} % bold math
\usepackage{hyperref}
\usepackage{multirow}
\usepackage{xcolor}
\usepackage{adjustbox}
\usepackage{enumitem}
\usepackage{caption}
\usepackage{anyfontsize}
\usepackage{color}
\usepackage{fontawesome5}
\usepackage{dsfont}
\usepackage{dcolumn}
\usepackage{url}
\usepackage{orcidlink}
\usepackage{subcaption}
\usepackage{graphicx}
\usepackage{multirow}

\graphicspath{{./figs/}}

\allowdisplaybreaks

\newcommand{\Tr}{\mathrm{Tr}}

\newcolumntype{\.}{>{\global\let\currentrowstyle\relax}}

\newcolumntype{^}{>{\currentrowstyle}}

\makeatletter
\newcommand{\ecitem}[2]{%
\item[\textbf{#1}]%
  \phantomsection
  \protected@edef\@currentlabel{\unexpanded{\textbf{#1}}}%
  \label{#2}%
}
\makeatother

\newcommand{\beq}{\begin{equation}}
\newcommand{\eeq}{\end{equation}}
\newcommand{\beqs}{\begin{eqnarray}}
\newcommand{\eeqs}{\end{eqnarray}}

\newcommand{\orcidauthorOHNO}{0000-0003-1798-8222}
\newcommand{\orcidauthorHSIAO}{0000-0002-8522-5190}
\newcommand{\orcidauthorTOMIYA}{0000-0001-9374-3716}
\newcommand{\orcidauthorCHOI}{0000-0002-5438-5490}

\title{\boldmath A Machine Learning Approach for Lattice Gauge Fixing}
\ShortTitle{A Machine Learning Approach for Lattice Gauge Fixing}

\author*[a]{Ho Hsiao\,\orcidlink{\orcidauthorHSIAO}}
\author[a]{Benjamin J. Choi\,\orcidlink{\orcidauthorCHOI}}
\author[a]{Hiroshi Ohno\,\orcidlink{\orcidauthorOHNO}}
\author[b,c,d]{Akio Tomiya\,\orcidlink{\orcidauthorTOMIYA}}

\affiliation[a]{Center for Computational Sciences, 
University of Tsukuba,\\
1-1-1 Tennodai, Tsukuba, Ibaraki 305-8577, Japan}

\affiliation[b]{Department of Information and Mathematical Sciences,
Tokyo Woman’s Christian University,\\
2-6-1 Zempukuji, Suginami-ku, Tokyo 167-8585, Japan}

\affiliation[c]{RIKEN Center for Computational Science, \\ 7-1-26
Minatojima-minami-machi, Chuo-ku, Kobe 650-0047, Japan}

\affiliation[d]{Department of Physics, Kyoto University, Kitashirakawa,
Sakyo-ku, Kyoto 606-8502, Japan}

\emailAdd{hohsiao@ccs.tsukuba.ac.jp}

\abstract{Gauge fixing is an essential step in lattice QCD
  calculations, particularly for studying gauge-dependent observables.
  Traditional iterative algorithms are computationally expensive and
  often suffer from critical slowing down and scaling bottlenecks on
  large lattices.
  We present a novel machine learning framework for lattice gauge
  fixing, where Wilson lines are utilized to construct gauge
  transformation matrices within a convolutional neural network.
  The model parameters are optimized via backpropagation, and we
  introduce a hybrid strategy that combines a neural-network-based
  transformation with subsequent iterative methods.
  Preliminary tests on SU(3) gauge theory ensembles for Coulomb gauge
  demonstrate the potential of this approach to improve the efficiency
  of lattice gauge fixing.
  Furthermore, we show that the model exhibits lattice size
  transferability, where parameters optimized on smaller lattices
  remain effective for larger volumes without additional training.
  This framework provides a scalable path toward mitigating critical
  slowing down in high-precision gauge fixing.}

\FullConference{The 42nd International Symposium on Lattice Field
  Theory (LATTICE2025)\\
  2-8 November 2025\\
  Tata Institute of Fundamental Research, Mumbai, India\\}

\begin{document}
%--------------------------------------------

%% %% \makebox[0.95\textwidth][r]{\preprint{CTPU-PTC-24-35}}
%% \begin{center}
%% \includegraphics[width=\textwidth]{PoSlogo.pdf}
%% \end{center}

%--------------------------------------------
\maketitle
%--------------------------------------------

%--------------------------------------------
\section{Introduction}
\label{sec:intro}
%--------------------------------------------

Gauge fixing, applying gauge transformations until a given condition,
is a crucial but computationally demanding process in lattice quantum
chromodynamics (QCD) calculations, particularly for large lattices
where critical slowing down becomes a significant challenge.
While gauge invariance is a fundamental feature of lattice
formulations, gauge fixing is often necessary for studying
gauge-dependent quantities, such as quark and gluon
propagators~\cite{Mandula:1987rh}, defining renormalization
schemes~\cite{Maiani:1991az}, or extracting fundamental QCD Green's
functions.
Furthermore, smearing with a gauge-fixed configuration is shown to
further reduce the noise in mass extractions~\cite{CP-PACS:2002unz}.
Traditional iterative algorithms for gauge fixing, such as the Los
Alamos (LA) method~\cite{Gupta:1987zc} and Cornell
method~\cite{Davies:1987vs}, are known to suffer from inefficiencies
as lattice sizes increase.
In both cases, the updating procedure is purely local, so that
information propagates only between neighboring lattice sites at each
iteration.
Motivated by these limitations of purely iterative algorithms with
local updates, we introduce a novel machine-learning framework for
lattice gauge fixing in SU(N) gauge theories to Landau (Coulomb)
gauge.
Our approach leverages Wilson lines to construct gauge transformation
matrices and convolute the link variables through a neural network to
efficiently acquire long distance information.
We employs backpropagation to compute gradients to update the model
parameters.  For machine learning, mini-batch training is applied to
learn optimal parameters.
The optimal goal of our study is to replace the iterative procedure by
a single gauge transformation.
We utilize the software package,
\texttt{Gaugefield.jl}~\cite{Nagai:2024yaf}, to compute gradient from
the neural network and implement the traditional gauge fixing methods,
while \texttt{Lux.jl}~\cite{pal2023lux} is employed for machine
learning.

In our initial study, we focus on the Coulomb gauge.
It can be easily extended to Landau gauge, as formulated in the
following sections.
We investigate the performance of our approach with two models built by different neural network depths and lengths of Wilson lines.
Various training schemes are tested to realize the training stability.
Preliminary results in SU(3) gauge theory demonstrate the potential
of this method to improve both the accuracy and computational
efficiency of gauge fixing.
Future work will explore the performance of the model across varying
lattice sizes and its adaptability to different gauge configurations,
including topologically frozen conditions, paving the way for more
robust and scalable lattice QCD simulations.

%--------------------------------------------
\section{Iterative Approaches}
\label{sec:LASD}
%--------------------------------------------

We briefly outline the iterative gauge-fixing procedures employed in
this study.
Comprehensive technical details are available in the original
publications~\cite{Gupta:1987zc,Davies:1987vs} and in representative
applications~\cite{Suman:1993mg,CP-PACS:2002unz}.
In lattice calculations, gauge fixing is achieved by satisfying the
following condition under certain gauge transformation $g$:
\begin{equation}
  F[g] = \frac{1}{d_{\text{fix}}\,N_c\,N_{\text{vol}}} \mathrm{Re}
  \sum_{n}\sum_{\mu}^{d_{\rm fix}} \textrm{Tr} \,
  U_\mu^{g}(n),\;\text{ with }\;\frac{\delta F}{\delta g} =0 .
  \label{eq:trU}
\end{equation}
The gauge-fixing functional, $F[g]$, computes the sum of traces of
$g$-transformed link variables, $U_\mu^{g}(n)$, over the fixing
direction, $d_{\rm fix}$.
In this work, we are interested in Landau (Coulomb) gauge fixing, for
which $d_{\text{fix}}=4$ ($d_{\text{fix}}=3$).
For the LA method, one defines
\begin{equation}
  w(n) = \sum_{\mu}^{d_{\rm fix}} U_\mu (n) + U^\dagger_\mu
  (n-\hat{\mu})\,.
\end{equation}
%
%% and reformulates Eq.~(\ref{eq:trU}) into even and odd lattice sites,
%% \begin{equation}
%%     F[g] = \mathrm{Re} \frac{1}{2} \textrm{Tr} \sum_{n \in {\rm even,odd}} w^{g}(n)\,.
%% \end{equation}
%
The gauge transformation is conducted on even or odd sites by
$w(n)\rightarrow w^{g}(n) = g(n)w(n)$, which project $w(n)$ alone its
SU(2) group manifold that satisfy $\mathrm{Re} \textrm{Tr}~g(n)w(n)
\geq \mathrm{Re} \textrm{Tr}~w(n)$.
This maximization is know as the Cabibbo-Marinari
method~\cite{Cabibbo:1982zn}, which is widely applied in the heat bath
algorithm.
The Cornell method essentially bases on the steepest-descent (SD)
approach, where the gauge transformation is given by
\begin{equation}
  g(n) = \exp{\alpha \Delta(n)}\,, \textrm{ where }\, \Delta(n) =
  \sum_{\mu} \left [U^\dagger_\mu (n-\mu) - U_\mu (n) \right
  ]_{\textrm{TA}}\,.
\end{equation}
The traceless anti-Hermitian (TA) operation follows the definition in Ref.~\cite{Nagai:2024yaf}.
The parameter $\alpha$, acting as the step size in SD, should be
properly chosen, either globally or locally, to guarantee efficient
convergence.
Hence, we further adopt the Mino method (see Appendix A in
Ref.~\cite{CP-PACS:2002unz}), that computes the local $\alpha$ as
\begin{equation}
  \alpha(n) = \frac{\displaystyle \mathrm{Re} \sum_{\mu} \textrm{Tr}
    \left [ \Delta(n) \left ( U_\mu(n-\hat{\mu}) - U_\mu (n)) \right )
      \right ] }{\displaystyle \mathrm{Re} \sum_{\mu}\textrm{Tr} \left
    [ \Delta^2(n) \left ( U_\mu(n-\hat{\mu}) + U_\mu (n) ) \right )
      \right ] } \,.
\end{equation}
An over-relaxation (OR) is also applied with a parameter $1<\omega<2$ in the gauge transformation by taking $g(n) = \exp{[\omega \alpha(n) \Delta(n)]}$.\footnote{
It is also possible to improve the LA method by introducing OR~\cite{Mandula:1990vs}. We only apply OR in the SD method for simplicity.
Furthermore, advanced techniques, such as Fourier acceleration~\cite{Davies:1987vs,Cucchieri:1998ew}, conjugate gradient~\cite{Hudspith:2014oja}, and global optimization~\cite{Oliveira:2003wa}, are beyond the scope of this study.
}
In practice, we combine these two method by first applying the LA
method with a given number of iterations to make sure monotonically
approaching gauge-fixed point, the maximum of Eq.~(\ref{eq:trU}).
Subsequently, we switch to SD, using the Mino method and an OR parameter $\omega=1.99$, until reaching desired tolerance 
with respect to $\Delta F[g]$, the change of $F[g]$.

%--------------------------------------------
\section{Model for Machine Learning}
\label{sec:model}
%--------------------------------------------

Our approach aims to bypass traditional iterative procedures by,
first, generalizing from local link variables to a sum of multi-length
Wilson lines, to capture long-range correlations.
Furthermore, we employ a convolutional neural network (CNN) to
effectively construct complicated Wilson lines in deep layers.
We then regard the SU(3) group projection as a non-linear activation
function within a machine learning architecture.
Lastly, machine learning optimization is applied to refine the
parameters of each length of Wilson lines at each layer.

For each layer, $\ell$, of the CNN, we define the Wilson line bundle
(WLB) gauge transform matrix at the lattice position $n$ as
\begin{equation}
  g^{\left(\ell\right)}\left(n\right) \equiv \exp\left[
    \Omega^{\left(\ell\right)}\left(n\right) \right]_{\text{TA}}\,,
  \textrm{ where }\, \Omega^{\left(\ell\right)}\left(n\right) =
  \sum_{r=1}^{s} \theta_{r}^{\left(\ell\right)} \;
  L_{r}\left(n;\left\{ U^{\left(\ell-1\right)}_\mu \right\} \right)
  \,.
\end{equation}
In this expression, $L_r(n;\{U^{(\ell-1)}_\mu\})$ denotes the
combination of length-$r$ Wilson lines with one end fixed to reference point $n$, which is constructed by
multiplying the gauge links of the previous layer, $U^{(\ell-1)}_\mu$.
The training parameters, $\theta^{(\ell)}_{r}$,
%% for $L_r$ of the layer $\ell$,
control the weight of such an object that contributes to the sum in
$\Omega^{(\ell)}(n)$ up to a desired maximum length $S$.
Since $\Omega^{(\ell)}(n)$ is a sum of SU(3) matrices, we take the
exponential function with traceless anti-Hermitian (TA) operation to
project it back to the SU(3) group for $g^{(\ell)}(n)$.
In the CNN construction, we assign the first layer as the original
link variables, namely, $U^{(1)}_\mu(n)\equiv U_\mu(n)$.
The link variables of the next layer is obtained by applying the WLB
gauge transformation as $ U^{(\ell)}_\mu = g^{(\ell)}(n) U^{(\ell
  -1)}_\mu(n) g^{(\ell) \dagger}(n+\hat{\mu}) \,.  $
In this way, we evolve the CNN to a number of layers $L$.
Notice that the effective number of training parameters $N_{\rm
  param.} = (L-1) \times S$, excluding the first layer with the
original gauge links.

In this model, the link variables at the final layer,
$U_\mu^{(L)}(n)$, is the output of the CNN, which optimally is
considered as a gauge-fixed configuration.
Hence, we compute the objective function defined by
\begin{equation}
  \mathscr{F}(\theta) = \frac{1}{d_{\text{fix}}\,N_c\,N_{\text{vol}}} \mathrm{Re}
  \sum_n \sum_{\mu=1}^{d_{\text{fix}}} \textrm{Tr} \, U_\mu^{(L)}(n),
  \label{eq:obj}
\end{equation}
where $d_{\text{fix}}$ denotes the number of directions depending the
gauge to fix, $N_c$ is the number of colors, and $N_{\text{vol}}$ is
the lattice volume.
The training procedure is formulated as the maximization of the objective
function.
We compute the gradient of Eq.~(\ref{eq:obj}) with respect to the
trainable parameters $\theta_r^{(\ell)}$ through backpropagation,
which takes the form:
\begin{equation}
  \frac{d\mathscr{F}(\theta)}{d\theta_r^{(\ell)}} = 2
  \sum_{\mu=1}^{d_{\text{fix}}} \sum_n \mathrm{Re}\,\Tr \left[
    \Lambda_\mu^{H(\ell)}(n)\, L_r\!\left(n;\{U^{(\ell-1)}\}\right) +
    \Lambda_\mu^{S(\ell)}(n+\hat\mu)\,
    L_r\!\left(n+\hat\mu;\{U^{(\ell-1)}\}\right) \right]\,,
\end{equation}
where
\begin{align}
  \Lambda_{\nu}^{H\left(\ell+1\right)}\left(m\right) & = \left[\left\{
    U_{\nu}^{\left(\ell\right)}\left(m\right)
    g^{\left(\ell+1\right)\dagger}\left(m+\hat{\nu}\right)
    \delta_{\nu}^{\left(\ell+1\right)}\left(m\right)\right\}
    \star\frac{\displaystyle \partial
      g^{\left(\ell+1\right)}\left(m\right)}{ \displaystyle \partial
      Q^{\left(\ell+1\right)}\left(m\right)}\right ]_{\text{TA}}\,,\\
  \Lambda_{\nu}^{S\left(\ell+1\right)}\left(m+\hat{\nu}\right) &
  =\left[\left\{ U_{\nu}^{\left(\ell\right)\dagger}\left(m\right) \;
    g^{\left(\ell+1\right)\dagger}\left(m\right) \;
    \delta_{\nu}^{\left(\ell+1\right)\dagger}\left(m\right) \right\}
    \star \frac{\displaystyle \partial
      g^{\left(\ell+1\right)}\left(m+\hat{\nu}\right) }{ \displaystyle
      \partial Q^{\left(\ell+1\right)}\left(m+\hat{\nu}\right)}
    \right]_{\text{TA}}\,.
\end{align}
The back-propagated contributions from the final layer,
$\delta_{\nu}^{\left(\ell\right)}\left(n\right)$ and
$\delta_{\nu}^{\left(\ell\right)\dagger}\left(n\right)$, are obtained
by
\begin{equation}
  \delta_{\mu}^{\left(\ell\right)}\left(n\right) =\frac{\displaystyle
    \partial \mathscr{F}\left(\theta\right)}{\displaystyle \partial
    U_{\nu}^{\left(\ell\right)}\left(n\right)} ~\text{ and }~
  \delta_{\mu}^{\left(\ell\right)\dagger}\left(n\right)
  =\frac{\displaystyle \partial \mathscr{F}\left(\theta\right)}{\displaystyle
    \partial U_{\nu}^{\left(\ell\right)\dagger}\left(n\right)} \,.
\end{equation}
We define the star product for the rank-2 and rank-4 tensors as
$
  [A \star T]^i{}_j \equiv \sum_{kl} A^l{}_k T^k{}_j{}^i{}_l.
$
The derivation of this backpropagation formula closely follows the
case of stout smearing in Ref.~\cite{Nagai:2021bhh}, and the detailed
derivation is deferred to a future publication.

%--------------------------------------------
\section{Training schemes}
\label{sec:training}
%--------------------------------------------

\begin{table}
  \vspace{-1.0em}
  \caption{JLDG public ensembles used in this work.  The ensembles
    were generated by the PACS-CS collaboration with the Iwasaki gauge
    action and the (2+1)-flavor non-perturbatively $O(a)$-improved
    Wilson quarks~\cite{PACS-CS:2008bkb}.
    For each ensemble, the spacial extent, $N_s$, the temporal extent,
    $N_t$, the lattice bare coupling, $\beta$, the clover coefficient
    $c_{\rm SW}$, the hopping parameter for u and d quarks,
    $\kappa_{\rm ud}$, the hopping parameter for s quark, $\kappa_s$,
    the lattice spacing, $a$, and the number of the available
    configurations, $N_{\rm conf.}$ are listed.}
%
%    These configurations are generated by the PACS-CS
%    collaboration~\cite{PACS-CS:2008bkb}.}
    \label{tab:ens}
    \centering
   \begin{tabular}{|c|c|c|c|c|c|c|c|c|}
    \hline\hline
    Ensemble & $N_{s}$ & $N_{t}$ & $\beta$ & $c_{\rm SW}$ & $\kappa_{\rm ud}$ & $\kappa_{\rm s}$  & $a{\rm [fm]}$ & $N_{\rm conf.}$ \\
    \hline\hline
    RC32x48 & 32 & \multirow{2}{*}{48} & \multirow{2}{*}{1.9} & \multirow{2}{*}{1.715} & \multirow{2}{*}{0.1373316} & \multirow{2}{*}{0.1367526} & \multirow{2}{*}{0.0907(13)} & 402 \\
    RC48x48 & 48 & & & &  & & & 200 \\
    \hline\hline
    \end{tabular}
    %\begin{tabular}{|c|c|c|c|c|c|c|c|}
    %\hline\hline
    %Ensemble & $N_{s}$ & $N_{t}$ & $\beta$ & $\kappa_{\rm ud}$ & $\kappa_{\rm s}$  & %$a{\rm [fm]}$ & $N_{\rm conf.}$ \\
    %\hline\hline
    %RC32x48 & $32 $ & $ 48$ & 1.9 & 0.1373316 & 0.1367526 & $0.0907(13)$ & 402 \\
    %RC48x48 & $48 $ & $ 48$ & 1.9 & 0.1373316 & 0.1367526 & $0.0907(13)$ & 200 \\
    %\hline\hline
    %\end{tabular}
\end{table}

To train our models, we make use of the JLDG public
ensemble~\cite{PACS-CS:2008bkb}, RC32x48, listed in
Table~\ref{tab:ens}, while the ensemble RC48x48 plays the role of a
validation set.
As this is our first attempt, we restrict ourselves to ensembles with
identical lattice parameters, differing only in the lattice spacial
extent.
We construct two CNN architectures, denoted as L21S2 and L12S3.
The former consists of 21 layers and incorporates Wilson-line
combinations of length up to two, and the latter has 12 layers and
extends the construction to including length-three Wilson lines.
We apply the mini-batch training, which is a widely used strategy that
strikes a balance between efficiency and stability during optimization
in machine learning.
Instead of updating model parameters using the entire training set
(full-batch training) or a single data point at a time (stochastic
training), mini-batch training processes small subsets of the data in
each iteration.
The training parameters are updated by the average of the gradients
computed from the configurations within a mini batch.
We employ adaptive moment estimation (Adam) optimizer to update the
model parameters with a learning rate $\alpha$.
A single training epoch is completed once all mini-batches within the
training set have been processed sequentially.

\begin{table}[b]
  \caption{Hyperparameters and the final results for each training
    scheme with ensemble RC32x48.
    The table details the training set size, mini batch size, CNN
    layer depth (L), maximum Wilson line length (S), the
    initialization strategy, and learning rate ($\alpha$).
    The final training objective value is reported alongside the total epochs.
    In the case of L12S3-4N-W, the notation "14+34" indicates that
    training is initialized from the 14th epoch of L12S3-1N-Z and
    continue for an additional 34 epochs.}
  \label{tab:skm}
  \centering
  \begin{tabular}{|c|c|c|c|c|c|c|c|c|}
    \hline\hline
    Scheme & Training set &  Batch size & L & S  & Initial param. & $\alpha$ & Epoch & $\mathscr{F}$ \\
    \hline\hline
    L21S2-1N-Z & 40 & 10 & 21 & 2 & zeros & 0.01 & 93 & 0.8576\\
    L21S2-4N-Z & 160 & 16 & 21 & 2 & zeros & 0.01 & 40 & 0.8576 \\
    \hline
    L12S3-1N-Z& 40 & 10 & 12 & 3  & zeros & 0.005 & 32 & 0.8498 \\
    L12S3-4N-Z & 160 & 16 & 12 & 3 & zeros & 0.005 & 39 & 0.8538\\
    L12S3-4N-W & 160 & 16 & 12 & 3 & L12S3-1N-Z & 0.005  & 14+34 & 0.8538\\
    \hline\hline
  \end{tabular}
\end{table}

In this work, we consider two sizes for the training set: a baseline
set of $40$ configurations processed with a mini-batch size of $10$,
denoted by the suffix "-1N", and a lager training set of $160$
configurations, denoted by "-4N", where the mini-batch size is $16$.
Most schemes are initialized from scratch with zero parameters
(indicated by the suffix "-Z"), where the gauge transformation matrix
is an identity matrix.
It allows us to investigate the effect on the size of the training
set.
Alternatively, we also perform an incremental training test
specifically with the L12S3 architecture.
In the scheme L12S3-4N-W, the training begins with the parameters
obtained at the $14$th epoch of the smaller training set.
The model then switches to the larger training set size and continues
for additional $34$ epochs.
This two-stage process is denoted by the epoch count $14+34$, and the
suffix, "-W", stands for a warm start.
The hyperparameters for each training scheme are summarized in
Table~\ref{tab:skm}, including the training set size, batch size, the
CNN architecture, initial training parameters, learning rate
$\alpha$, the number of conducted epoch, and the final value of the
objective function.

\begin{figure}
  \vspace{-1.5em}
  \centering
  \begin{subfigure}[t]{\textwidth}
    \centering
    \includegraphics[width=\linewidth]{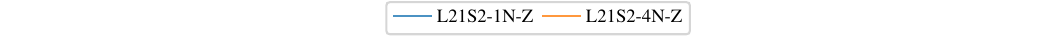}
  \end{subfigure}
  \begin{subfigure}[t]{0.6\textwidth}
    \centering 
    \includegraphics[width=\linewidth]{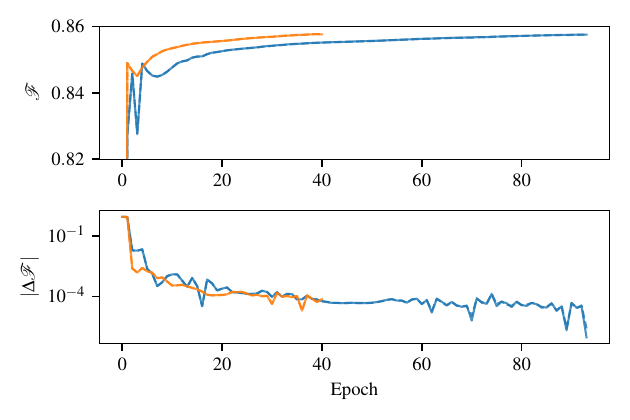}
    \subcaption{The objective function and its absolute difference between two epochs.}
  \end{subfigure}
  \begin{subfigure}[t]{0.39\textwidth}
    \centering
    \includegraphics[width=\linewidth]{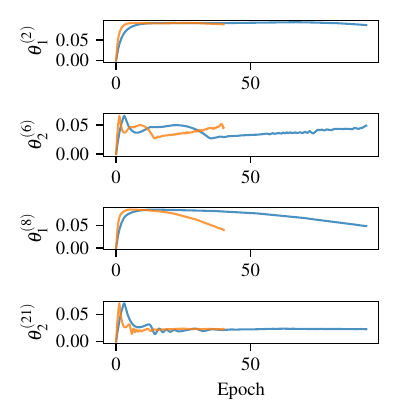}
    \subcaption{Evolution of selected training parameters.}
  \end{subfigure}
  \caption{Training history of the L21S2 CNN, consisting of 21 layers
    with Wilson lines of lengths one and two.
    The blue curves represent the L21S2-1N-Z scheme (small dataset),
    and the orange curves denote the L21S2-4N-Z scheme (large
    dataset).}
  \label{fig:skm_L21S2}
\end{figure}

\begin{figure}
  \vspace{-1.5em}
  \centering
  \begin{subfigure}[t]{\textwidth}
    \centering
    \includegraphics[width=\linewidth]{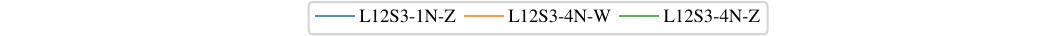}
  \end{subfigure}
  \begin{subfigure}[t]{0.6\textwidth}
    \centering 
    \includegraphics[width=\linewidth]{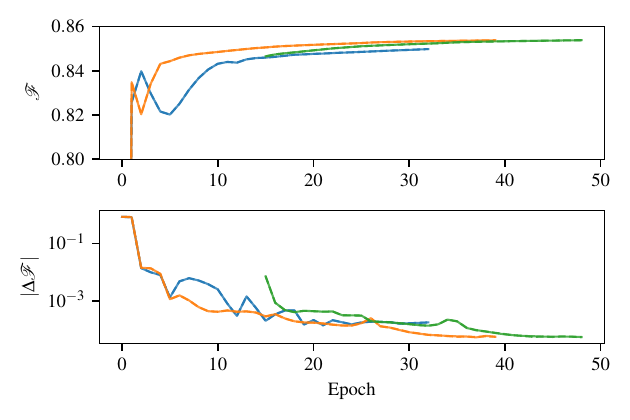}
    \subcaption{The objective function and its absolute difference between two epochs.}
  \end{subfigure}
  \begin{subfigure}[t]{0.39\textwidth}
    \centering
    \includegraphics[width=\linewidth]{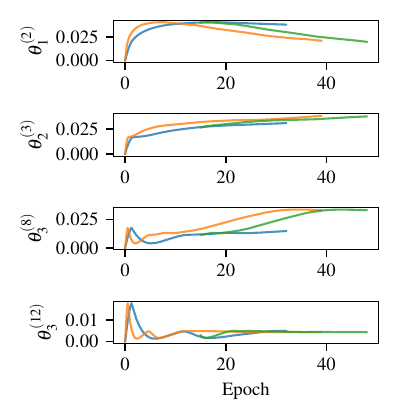}
    \subcaption{Evolution of selected training parameters.}
  \end{subfigure}
  \caption{Training history of the L12S3 CNN, featuring 12 layers
    and Wilson lines up to length three.
    The blue curves denote the L12S3-1N-Z scheme (small dataset),
    while the orange curves represent L12S3-4N-Z (large dataset).
    The green curves illustrate the incremental training scheme
    (L12S3-4N-W), where training resumed from the 14th epoch of the
    small-set run using the larger dataset.
    Its large value of $|\Delta \mathscr{F}|$ at the beginning
    reflects the reconfiguration of the model as it adapts to the
    richer dataset, eventually aligning with L12S3-4N-Z.}
  \label{fig:skm_L12S3}
\end{figure}

Figures~\ref{fig:skm_L21S2} and~\ref{fig:skm_L12S3} show the training
history with respect to the epoch for CNN architectures L21S2, and
L12S3, respectively, where the legend denotes the training scheme.
In the left panels, we present the objective functions, ${\mathscr F}$, defined in
Eq.~(\ref{eq:obj}) (top) and their relative change comparing to that
of the previous epoch (bottom), denoted as $|\Delta {\mathscr F}|$.
In the right panels, selective training parameters,
$\theta^{(\ell)}_r$, for Wilson line with a length $r$ at layer
$\ell$, are displayed.
On one hand, as shown in Figure~\ref{fig:skm_L21S2}, the training for
the L21S2 architecture reaches to the same final result for both the
small (blue curves) and large (orange curves) datasets, in terms of
the objective function and parameter values.
On the other hand, the L12S3 architecture appears more sensitive to
dataset size.
As illustrated in Figure~\ref{fig:skm_L12S3}, $\Delta {\mathscr F}$ for the smaller dataset (blue) significantly
plateaus after approximately 25 epochs.
A comparison between L12S3-1N-Z (blue) and L12S3-4N-Z (orange), both
starting from zero parameters, reveals significant
discrepancies not only in the final objective values but also in the
parameters, especially $\theta_1^{(2)}$ and $\theta_3^{(8)}$.
This suggests that the smaller dataset lacks the necessary information
to capture the complexity of the model.
\footnote{Although the number of training parameters for L21S2
($N_{\rm param.} = 40$) is larger than that for L12S3 ($N_{\rm param.}
= 33$), the complexity of a model is driven by the combinations of
involving Wilson lines.
In the case of $d_{\rm fix} =3$, the number of combinations with the
length-$2$ Wilson lines is $30$, while that of the length-$3$ lines is
$150$.
This necessitates a larger training set for proper generalization.}
To explore potential optimizations for future large-scale runs, we
introduced an incremental training strategy in scheme L12S3-4N-W
(green) by resuming training from the 14th epoch of the small-set run
using the larger dataset.
This transition causes a momentary increase in the objective function,
followed by a correction that aligns the parameters and objective value with the
L12S3-4N-Z results.
This confirms that even when a model begins to converge toward a
suboptimal steady-state due to limited data, an incremental shift to a
larger dataset can successfully steer the parameters toward the
behavior observed in the training with a larger set.

%--------------------------------------------
\section{Preliminary results for gauge fixing}
\label{sec:result}
%--------------------------------------------

In this section, we present preliminary results for the gauge-fixing
performance of our trained models, using outcomes from traditional
methods as a baseline for comparison.
As shown in Figures~\ref{fig:skm_L21S2} and~\ref{fig:skm_L12S3}, the
objective functions increase during the trainings, which are expected to
approach the value evaluated from $F[g]$ using the traditional
iterative approach.
Given the limited training duration of these exploratory runs, the
final objective values have not yet reached the theoretical optimal value
of $0.86965(8)$; see Table~\ref{tab:skm} for final values.
However, the primary goal of this contribution is not to achieve
immediate convergence, but rather to explore the potential of the
model to accelerate the gauge-fixing process.
Specifically, we investigate whether the hybrid procedure, applying
first WLB gauge transformation then the standard iterative approach,
can significantly reduce the total number of iterations required to
reach the target precision.
%
%% After the WLB gauge transformation,
%
To ensure a fair comparison, the number of iterations in the LA method
is chosen to yield an $F[g]$ value equivalent to that of the pure
iterative approach.
The convergence tolerance for the subsequent SD method is set to
$\Delta F[g] < 10^{-12}$.
We select two representative sets of training parameters from
L21S2-1N-Z and L12S3-4N-W to conduct such validation tests.

\begin{figure}
  \vspace{-2.0em}
  \centering
  \includegraphics[width=\linewidth]{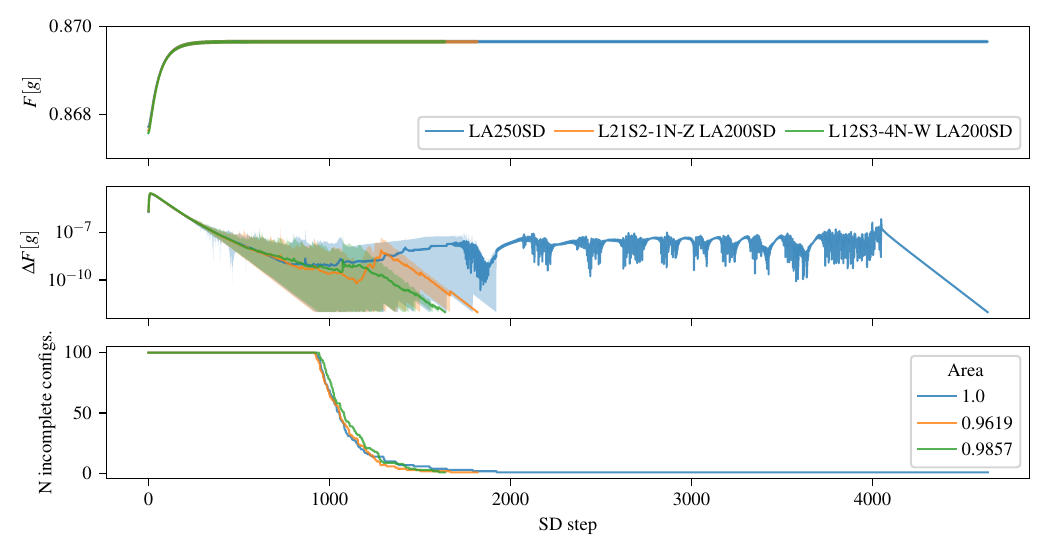}
  \caption{Gauge-fixing performance for the RC32x48 ensemble across
    100 test configurations.
    Comparisons are shown between the pure iterative baseline (blue,
    beginning with 250 LA steps) and hybrid schemes using L21S2-1N-Z
    (orange) and L12S4-4N-W (green) trained parameter.
    The panels display: (Top) Evolution of the gauge-fixing functional
    $F[g]$, with solid lines indicating mean values and bands
    representing standard deviations.
    (Middle) Relative difference $\Delta F[g]$ between successive
    iterations; bands denote the range between maximum and minimum
    values across incomplete configurations.
    (Bottom) The number of incomplete configurations remaining
    throughout the SD iterations.
    Normalized total computational costs are provided in the legend,
    representing the area under each curve relative to the pure
    iterative baseline.}
  \label{fig:fixing_RC32x48}
\end{figure}

Figure~\ref{fig:fixing_RC32x48} demonstrates the results of gauge
fixing ensemble RC32x48, with $100$ configurations that are not
involved in trainings.
The blue curves denote the results of the pure iterative approach,
with first $250$ iterations in the LA method.
The orange and green curves represent the outcomes from L21S2-1N-Z and
L12S4-4N-W, respectively, with only $200$ LA iterations ($50$
iterations reduced).
The top panel display the gauge-fixing functional, $F[g]$, with
respect to the iteration step in SD method, where solid lines are the
mean values, and the width of bands correspond to standard deviations.
In the middle panel, we illustrate the relative difference between two
iterations.
The solid lines are the average, while the upper (lower) bounds of the
bands are the maximum (minimum) over the incomplete configurations.
We present the number of the incomplete configurations across the SD
iterations in the bottom panel.
The area below each curve corresponds to the total computational cost
to fix all the configurations, which is normalized by such quantity
obtained from the pure iterative approach as shown in the legend.

\begin{figure}
  \vspace{-2.0em}
  \centering
  \includegraphics[width=\linewidth]{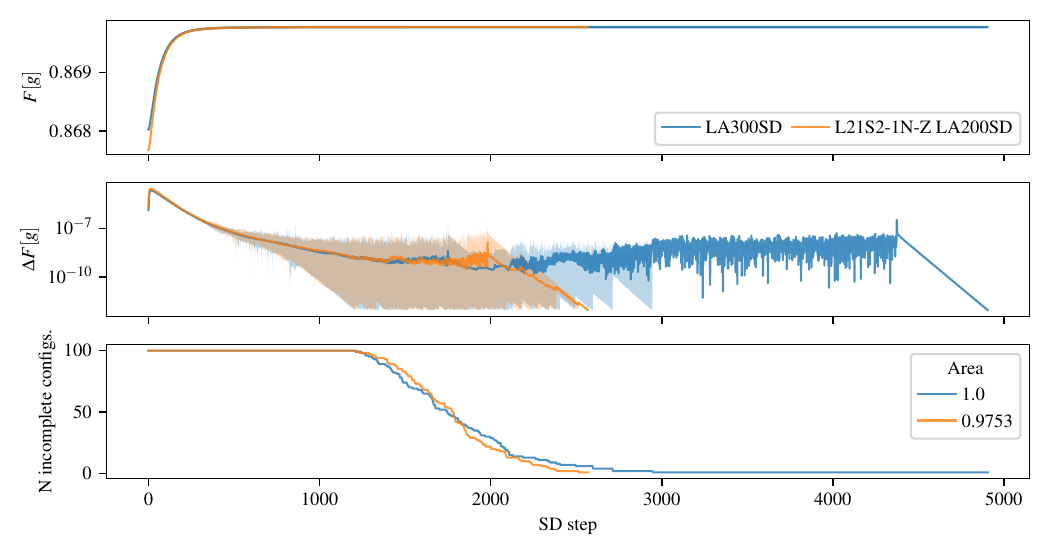}
  \caption{Gauge-fixing performance for the RC48x48 ensemble across
    100 test configurations.
    The plot compares the pure iterative baseline (blue, starting with
    300 LA steps) against the hybrid approach with trained parameters
    from L21S2-1N-Z scheme (orange).
    The panels illustrate: (Top) The evolution of the gauge-fixing
    functional $F[g]$, where solid lines represent mean values and
    shaded bands indicate standard deviations.
    (Middle) The relative difference $\Delta F[g]$ between successive
    iterations; the bands capture the range between the maximum and
    minimum values among incomplete configurations.
    (Bottom) The count of remaining incomplete configurations as a
    function of SD iterations.
    As indicated in the legend, the hybrid approach achieves a
    normalized total computational cost of 0.9753 relative to the pure
    iterative baseline. }
  \label{fig:fixing_RC48x48}
\end{figure}

Although the neural network models were trained for a limited
duration, the hybrid strategy demonstrates a clear advantage over the
pure iterative baseline.
As shown in the top panel, $F[g]$ for all schemes converges toward the
same asymptotic limit.
The improvement is evidenced in the middle panel, where the relative
difference $\Delta F[g]$ in both orange and green decreases
consistently, indicating that the initial WLB gauge transformation
provides a configuration that is closer to the gauge fixed point.
This smooth convergence avoids the critical slowing down seen in the
pure iterative approach (blue).
The total computational effort (bottom panel) yields efficiency
factors of 0.9619 for the L21S2-1N-Z scheme (orange) and 0.9857 for
the L12S3-4N-W scheme (green).
These reductions also reflect the fact that L21S2-1N-Z acquires a
larger objective value, compared to L12S3-4N-W; see
Table~\ref{tab:skm}.
Based on the superior performance and higher final functional value
achieved by the L21S2-1N-Z scheme, we selected this specific model to
investigate the lattice-size transferability of our trained parameters
to a larger lattice volume, ensemble RC48x48, without further
training.
In the top panel of Figure~\ref{fig:fixing_RC48x48}, the hybrid
approach (orange) reaches to the same value as the traditional one
(blue).
The middle and bottom panels confirm that the relative difference
$\Delta F[g]$ decreases consistently and the count of incomplete
configurations drops steadily, resulting in a normalized computational
cost of 0.9753.
This demonstrates that the local gauge features learned by the L21S2
architecture are not only effective but also highly scalable, allowing
the model to be trained on smaller, less expensive volumes while
remaining valid for large-scale production runs.

%--------------------------------------------
\section{Summary and Outlook}
\label{sec:summary}
%--------------------------------------------

In this work, we introduced a machine learning framework for lattice
gauge fixing designed to mitigate the computational bottlenecks and
critical slowing down characteristic of traditional iterative methods.
By utilizing Wilson lines within a CNN to construct gauge
transformation matrices, we successfully implemented a model capable
of maximizing the gauge-fixing functional for Coulomb gauge.
Our hybrid strategy demonstrates a clear improvement in efficiency
over the pure iterative approach, providing smooth and stable
convergence throughout the process.

A significant finding of this study is the volume transferability of
the trained parameters.
Our model, processing translational invariant property, trained on a
smaller lattice can apply successfully to a larger volume without
further training.
This demonstrates that the local gauge structures captured by the CNN
are volume-independent, providing a scalable principle for future
optimizations.
Furthermore, we verified that an incremental training approach, moving
from small to large datasets, allows the model to reconfigure its
parameters to adapt to richer information density while reducing
overall training cost.

In the ongoing research, we plan to evaluate the performance of this
framework across even larger lattice sizes and investigate its
adaptability to topologically frozen configurations.
These results provide more robust and scalable applications, with the
ultimate goal of replacing iterative procedures with a single, highly
efficient gauge transformation.

%--------------------------------------------
\acknowledgments

The authors thank the members of the project "Search for physics
beyond the standard model using large-scale lattice QCD simulation and
development of AI technology toward next-generation lattice QCD",
especially Takeshi Yamazaki, for useful discussions.
H. H., H. O., and A. T., were supported by JSPS KAKENHI Grant
No.~22H05112.
The work of A. T. was partially supported by JSPS KAKENHI Grants
No.~20K14479, No.~22H05111, No.~22K03539 and JST BOOST, Japan Grant
No.~JPMJBY24F1.
B. J. C. and part of this work were supported by MEXT as ``Program
for Promoting Researches on the Supercomputer Fugaku'' (Grant Number
JPMXP1020230411, JPMXP1020230409).
This research used computational resources of Pegasus and Miyabi
provided by Multidisciplinary Cooperative Research Program in Center
for Computational Sciences, University of Tsukuba.
%--------------------------------------------

%--------------------------------------------
%--------------
% bibliography
%--------------
\bibliography{bibliography}
%--------------------------------------------

%--------------------------------------------
\end{document}